\documentclass[aps,pre,twocolumn,groupedaddress]{revtex4}
\usepackage{amsmath}
\usepackage{graphicx}
\usepackage[mtbold,uprightgreek,mtpluscal,noTS1]{}

\begin{document}

\title{Computational analysis of folding and mutation properties of 
       C5 domain from Myosin binding protein C}

\author{Carlo Guardiani} 
\affiliation{Centro Interdipartimentale per lo Studio 
delle Dinamiche Complesse (CSDC) \\
Sezione INFN di Firenze, (Italy)} 

\author{Fabio Cecconi}
\affiliation{INFM-SMC and Istituto dei Sistemi Complessi ISC-CNR, \\
         Rome Italy}

\author{Roberto Livi}
\affiliation{Dipartimento di Fisica Universit\`a di Firenze \\
Centro Interdipartimentale per lo Studio 
delle Dinamiche Complesse (CSDC),\\
Sezione INFN di Firenze e INFM UdR Firenze, Italy.}

\begin{abstract}
Thermal folding Molecular Dynamics simulations of the domain C5 from
Myosin Binding Protein C were performed using a native-centric
model to study the role of three mutations related to 
Familial Hypertrophic Cardiomyopathy. Mutation of Asn755
 causes the largest shift of the folding temperature, and 
the residue is located in the CFGA' $\beta$-sheet featuring 
the highest $\Phi$-values.
The mutation thus appears to reduce the thermodynamic stability 
 in agreement with experimental data. The mutations on Arg654 and
Arg668, conversely, cause a little change in the
folding temperature and they reside in the low $\Phi$-value
BDE $\beta$-sheet, so that their pathologic role cannot be
related to impairment of the folding process but possibly to 
the binding with target molecules. As the typical signature of 
Domain C5 is the presence of a longer and destabilizing CD-loop with 
respect to the other Ig-like domains we completed the work
with a bioinformatic analysis of this loop showing a 
high density of negative charge and low hydrophobicity. This 
indicates the CD-loop as a \emph{natively unfolded sequence} with a 
likely coupling between folding and ligand binding.
\end{abstract}

\maketitle

{\bf Keywords:}
Simulations, Myosin binding protein C,  Folding,  Mutations, Natively Unfolded
Proteins;  

\section{Introduction}
\label{Intro}
\begin{figure}
\begin{center}
\includegraphics[clip=true,keepaspectratio,scale=0.2]{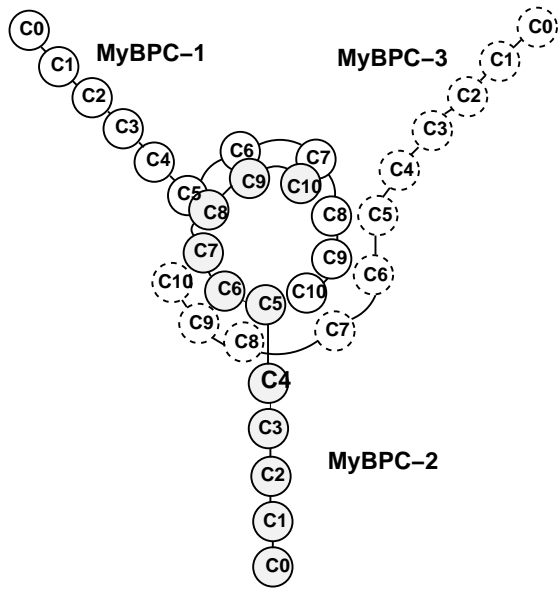}
\end{center}
\caption{Moolman-Smook model of MyBP-C arrangement in the sarcomere:
         the trimeric collar surrounding the thick filament is
         stabilized by interactions between domains C5-C8 and C7-C10.}
\label{ms-model}
\end{figure}

Familial Hypertrophic Cardiomyopathy (FHC) is a  genetic disease causing 
significant impairment of cardiac functionality and premature death in 
children and young adults \cite{Winegrad}. 
A number of mutations in genes encoding cardiac sarcomeric proteins
including the $\beta$-myosin heavy chain, the cardiac troponin T, 
titin, and cardiac myosin binding protein C (MyBP-C) have been found 
to correlate with such disease~\cite{Chung,Kimura,Mogen,Poetter,Satoh}. 
The FHC patients with MyBP-C mutations 
represent 20-45 \% of the total~\cite{Niimura, Charron}, so that 
mutations on this protein 
are the second most common  cause of the disease. 
While nonsense mutations on MyBP-C gene determine
a premature termination of translation of the C-terminus and result in a 
mild phenotype~\cite{Niimura, Charron,Charron2}, a number of missense mutations 
lead to a severe  phenotype and 
the precise mechanism through which they cause the disease is still 
unknown~\cite{Daehmlow,Yu}.      
MyBP-C is a linear sequence of 11 IgI-like and fibronectin-like domains
referred to as C0-C10 working as a potential regulator of cardiac 
contractility \cite{Winegrad}.
According to Moolman-Smook model \cite{Moolman,Review},
three MyBP-C molecules form a ring around the thick filaments 
(Figure~\ref{ms-model}). The collar
is stabilized by specific interactions between domains C5-C8 of a 
molecule and  domains C7-C10 of the neighboring one. The amino-terminal 
region between domains C0 and C4 protrudes out of the thick filament 
and, upon phosphorylation, interacts with subfragment 2 of Myosin (S2) 
acting as a brake for muscular contraction~\cite{Kunst}.  Mutations on MyBP-C 
might either prevent the C0-C4 region from interacting with the S2 fragment
(hyper-contractility) or force a carboxy-truncated mutant to permanently 
interact with S2 (hypo-contractility)~\cite{Review,Redwood}.

Our work will be concerned with the folding behavior of domain C5 whose 
structure was resolved through NMR by Idowu et al. \cite{domainC5}. This 
domain belongs to the IgI set of the Immunoglobulin superfamily and it 
features a typical $\beta$-sandwich structure with two twisted 
$\beta$-sheets closely packed against each other. The first $\beta$-sheet 
($\beta$1) is formed by strands C,F,G and A', while the second 
$\beta$-sheet  comprises strands B,D and E (see Fig.~\ref{fig:struct}).
A remarkable peculiarity of the C5 domain of the cardiac MyBP-C isoform 
is the presence of two long insertions not present in the fast and slow 
skeletal isoforms. The first insertion is 10-residue long and is located
in the linker between the C4 and C5 domains; the second insertion 
is 28 residues in length and resides in the CD loop~\cite{domainC5}.

A recent experimental work \cite{domainC5} proved that the N-terminal 
region containing the first insertion is not just a linker between C4 and C5, 
but it plays an important role in the thermodynamical stability
of the domain. 
Conversely, the long and highly mobile prolin-reach CD-loop  
destabilizes the protein lowering the folding temperature as compared to 
other Ig-domains\cite{domainC5}.  This loop is suspected  to form an SH3 domain recognition 
sequence presumably binding to the CaM-II like Kinase that co-purifies with 
MyBP-C~\cite{Oakley,Hartzell}. 
  
Three FHC causing mutations have been identified on C5 domain:
Asn755Lys, Arg654His and Arg668His. The first one lies on the FG-loop 
and leads to a significant destabilization of the protein yielding a 
severe phenotype~\cite{Moolman,domainC5,Singh}. 
The Arg654His and Arg668His mutations, related to a much milder phenotype, 
are reported not to impair the thermodynamic stability of the protein. 
Residue  Arg654 is actually suspected to regulate the specificity of the 
binding of positively charged
substrates, as it is located in the negatively charged CFGA' face that 
is a potential target for the binding with domain 
C8~\cite{Moolman,domainC5,Daehmlow}. A role in ligand 
binding is also postulated for Arg668~\cite{Morner}. 

The purpose of the present work is to investigate through MD simulations
the role of the above mentioned three known FHC causing mutations.
Clarke and coworkers showed that for the proteins belonging to the 
Ig-superfamily both transition and native states are stabilized by the same 
contacts dictated by protein topology~\cite{Clarke}.
This implies that the Ig-superfamily members share similar 
folding pathways mainly determined by their common geometry. Therefore
the native-centric approach seems to be the natural framework to 
investigate the folding properties of Ig-like C5 domain. 
To incorporate topology and some specific chemical feature as well as
the effect of the side chain packing we resorted to 
consider a heavy-map G\={o} model where native contacts
are identified on the basis of the steric hindrance of side-chains. 
Moreover we also introduced some amount of   
heterogeneity in the energetic couplings of the G\={o}  force field.
This approach is particularly suitable for the C5 domain 
because we need to address the problem of 
discriminating the effects of mutations such as Arg654His and Arg668His
modeled through the removal of the same number of native contacts.

The importance of introducing energetic heterogeneity in a G\={o}-model
for a reliable mutation analysis, is showed for example,
by the work of Clementi \emph{et al.}~\cite{ClementiGoTune}, where all the
available experimental data on free energy differences upon single mutations
of S6 ribosomal protein and its circular permutants, were reproduced with
correlation coefficients larger than $0.9$.

\section{Methods}
When the folding process is mainly driven by the topological constraints
of the native state, it is convenient to use simplified coarse-grained models 
describing the protein molecule as a chain of beads centered on the 
C$_{\alpha}$ carbon 
positions~\cite{Riddle,Chiti,Plaxco,Cecconi,Cecconi2,Cieplak}. 
The G\={o} force field introduces a bias 
towards native structure rewarding native-like interactions, through 
Lennard-Jones attractive forces and appropriate angular potentials 
embodying secondary structure motifs. The approach 
assigns to the native state the lowest energy and minimizes 
frustration yielding a perfect funnel landscape. 
We consider a light variant of the G\={o}-like force field proposed 
by Clementi \emph{et al.} and used in several other papers~\cite{Cecconi3,Kouza,BrooksIII}. 
The model is defined by the Potential Energy~\cite{Clementi}:
\begin{eqnarray}
V_{tot} = \sum_{i=1}^{N-1}\frac{k_h}{2}(r_{i,i+1} - R_{i,i+1})^2 + 
\sum_{i=2}^{N-1} \frac{1}{2}k_{\theta}(\theta_{i} - \theta_{i}^0)^{2} + 
\nonumber \\ 
\sum_{i=3}^{N-2} k_{\phi}^{(1)}[1 - \cos(\phi_{i} - \phi_{i}^{0})] + 
k_{\phi}^{(3)}[1 - \cos3(\phi_{i} - \phi_{i}^{0})] 
\nonumber \\
\sum_{i,j>i+3} V_{nb}(r_{ij})\;. 
\end{eqnarray}
Where the last potential corresponding to non bonded interaction is such 
that 
\begin{equation}
V_{nb}(r_{ij}) =
\left\{   
\begin{array}{ll} 
\epsilon_{ij} \left[5 \left(\frac{R_{ij}}{r_{ij}} \right)^{12} -
6 \left (\frac{R_{ij}}{r_{ij}} \right)^{10} \right]   & 
\mbox{$i-j$ native} \\ 
5 \epsilon_r \left(\frac{\sigma}{r_{ij}} \right)^{12} & 
\mbox{$i-j$ not native}
\end{array}
\right.
\end{equation}
In the above formulas, $r_{ij}$ is the distance between residue $i$ and $j$, 
$\theta_{i}$ 
is the bending angle identified by the three consecutive C$_{\alpha}$'s
$i-1$, $i$, $i+1$, $\phi_{i}$ is the dihedral angle defined by the two 
planes formed by four consecutive C$_{\alpha}$'s $i-2$, $i-1$, $i$, $i+1$.
The symbols with the superscript $0$ and $R_{ij}$ are the corresponding 
quantities in the native conformation. 
The force field parameters are
proportional to the energy scale $\epsilon_0 = 0.3$ Kcal/mol
such that $k_h = 1000 \epsilon_0/r_0^2$  ($r_0 = 3.8$ \AA),   
$k_{\theta} = 20\epsilon_0$,   
$k_{\phi}^{(1)} = \epsilon_0$ and  $k_{\phi}^{(3)} =0.5\epsilon_0$.
The parameters of the repulsive Lennard-Jones terms between non native 
contacts are chosen as follows: $\sigma= 5.0$~\AA, 
$\epsilon_r = 2/3 \epsilon_0$.
Two residues $i$ and $j$ are considered to interact attractively whenever 
their side-chains have at least a pair of
heavy atoms closer than a distance cutoff $R_c = 5$ \AA. Accordingly,  
the attractive native interactions depend on the coefficients
$\epsilon_{ij} = \epsilon_0 (1 + n_{ij}/n_{max})$, where $n_{ij}$ is the 
number of atomic contacts between residues $i$ and $j$ in their native 
position and $n_{max} = 43$ is the maximum value of $n_{ij}$ in the set
of native contacts corresponding to the couple Lys45-Tyr109. 
We performed controlled temperature MD simulations 
employing the isokinetic thermostat~\cite{isokin} with
time step $h = 5\times 10^{-3} \tau$,
where the time unit $\tau = \sigma\sqrt{M/\epsilon_0} = 4.67$ ps ($M$ is the
average mass of an aminoacid residue estimated to $110$ Da).

As a remark we observe that several strategies can be employed to introduce 
heterogeneity. A very common choice is to use the set of parameters derived 
by Miyazawa and Jernigan \cite{Miyazawa,Karanicolas}. Other 
authors~\cite{Clementi} instead prefer to tune the energy parameters
through a design procedure based on energy gap maximization.  
Our strategy of rescaling the contact energies according to the number 
of atomic contacts is grounded on experimental and theoretical evidence.
In particular, the mutation analysis on barnase by Serrano \emph{et al.}
\cite{Serrano} showed a non trivial correlation between the destabilization
induced by the mutation and the number of methyl or methylene side groups 
surrounding the deleted group. Moreover, Kurochkina and Lee~\cite{Kurochkina},
found that the pairwise sum of the buried surface area is linearly related
to the true buried area, as computed with the algorithm of Lee and Richards
\cite{Richards} and to the contact potential of Miyazawa and 
Jernigan~\cite{Miyazawa}. The approach proposed by Kurochkina was then 
followed by Sung~\cite{Sung} for an efficient modeling of the hydrophobic 
effect in a Monte Carlo study of $\beta$-hairpin folding. 
A significant correlation between
the average contribution of individual residues to folding stability and 
the buried ASA was also noticed by Zhou and Zhou~\cite{Zhou}.     
To check that heterogeneity we have introduces does not lead to excessive 
frustration in the energy landscape, we performed rapid quenching simulations 
to collect a data set $10^3$ of decoys. 
We then estimate the ratio $T_g/T_f$ between 
the glassy and folding temperature of our protein as the ratio: 
energetic standard deviation of the decoy set over the energy gap.   
We found that this quantity which is a measure of the energy landscape 
frustration remained substantially unaltered  
from the heterogeneous G\={o}-model, $T_g/T_f = 0.34$, to 
the homogeneous one $T_g/T_f = 0.35$. 
 
A customary indicator of the native-likeness of residues in the transition
state  (TS) is represented by the $\Phi$-values: 
a value $\Phi \sim 1$, characterizes residues establishing native-like
interactions already in the TS, whereas a value close to zero is typical 
of residues involved into a disordered conformation in the TS.
We apply the free energy perturbation technique (FEP) \cite{Clementi} 
to evaluate the $\Phi$-values from our MD simulations
\begin{equation}
\Phi =
\frac{
\log\langle\exp\{-\Delta E/RT \}\rangle_{TS} -
\log\langle\exp\{-\Delta E/RT \}\rangle_{U}}
{\log\langle\exp\{-\Delta E/RT \}\rangle_{F} -
\log\langle\exp\{-\Delta E/RT \}\rangle_{U}
}
\label{FEP_phi}
\end{equation}
where the Boltzmann factors depend on the energy difference between the
mutant and the wild type (WT) and the averages are computed over
WT-conformations of the folded (F), transition state (TS) and unfolded (U)
ensembles.
In the present paper, the $\Phi$-values are computed according to
equation~\ref{FEP_phi} using a method developed by Clementi 
{\em et al.} \cite{Clementi} that can be summarized in the following steps.
i) Determination of the folding temperature $T_f$ from the specific
heat plot. ii) Analysis of the free energy profile at temperature $T_f$ 
plotted as a function of a suitable folding reaction coordinate.
The double-well free energy profile of a two-state folder 
allows to define three windows of the reaction coordinate
identifying the folded, transition state and unfolded ensembles respectively.
iii)  Dynamic simulation at $T = T_f$ and storage of conformations
belonging to the F, TS and U ensembles. vi) 
Choice of mutations and computation of FEP $\Phi$-values (\ref{FEP_phi}).

Structural information about the native-likeness of the transition state
was also gained from the so-called structural $\Phi$-values:
\begin{equation}
\Phi_{struc}(i) =
\frac{\sum_{j\in C(i)}\;P_{TS}(i,j)}
{\sum_{j\in C(i)}P_{F}(i,j)}
\label{eq:Phistruct}
\end{equation}
where $P_{F}(i,j)$ and $P_{TS}(i,j)$ are the frequencies of the native
contact $i-j$ in the folded and
transition ensembles respectively, and the sum runs over the set
$C(i)$ of native contacts in which residue $i$ is involved.

An interesting property of the Transition State is the existence
of a few \emph{key residues} acting as nucleation centers for
the folding process. Following an approach proposed by
Vendruscolo \emph{et al.}~\cite{Vendruscolo}, the importance of 
these residues can be better understood by portraying the protein 
as a weighted graph. Residues represent the vertices 
and the weighted edges are defined as $w_{ij} = 1/A_{ij}$ where 
$A_{ij}$  represents  the fraction of TS ensemble structures 
where residues $i$ and $j$  are in contact. By using the Dijkstra's 
algorithm~\cite{Dijkstra} we computed the minimal path 
$\lambda_{ij}$ \emph{i.e.}
the minimum of the sum of the weights $w_{kl}$ of the edges traversed
along each route between $i$ and $j$. The fraction of minimal paths
passing through residue $k$ defines the \emph{betweenness} $B_k$ of
that residue. This quantity therefore measures the centrality 
of a residue: residues with a high betweenness act as "hubs" in the
network and they presumably  play a crucial role in the 
stabilization of the transition state.    
 
\section{Results}
A thermal folding simulation of the wild-type (WT) C5-domain was performed
by gradually cooling the protein from temperature $T = 2.5$ to
$T = 1.5$ in 50 temperature steps. For each temperature, an equilibration 
stage of $5\times10^{7}$ time steps was followed by a production stage
of $5\times10^{8}$ time steps. A similar schedule was employed to
simulate the folding of the three missense mutants Asn115Lys, Arg14His
and Arg28His (notice that protein residues have been renumbered 1-130 
as a restriction to the C5 domain only). Within the framework of the 
G\={o}-model, we decided to implement a mutation of a residue by 
turning all its native contacts into non-native ones. 
The role of the amino-terminal region of the protein was also investigated 
through folding simulations of 
a deletion mutant, where the first 7 residues of the C5 domain were removed. 

\begin{figure}
\begin{center}
\includegraphics[clip=true,keepaspectratio,scale=0.35]{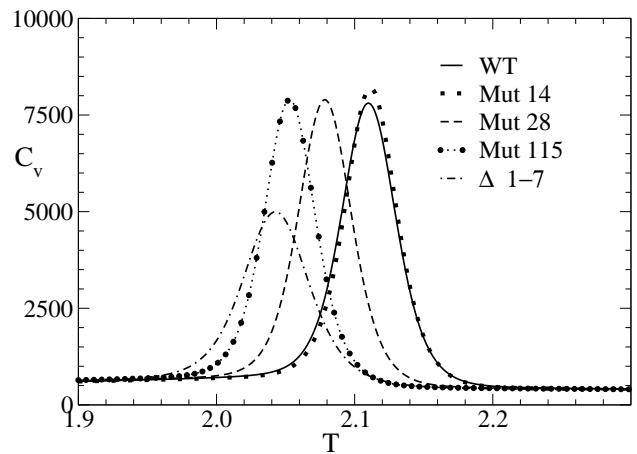}
\end{center}
\caption{Thermal behavior of heat capacity of the WT C5 domain,
         the missense mutants deprived of the native contacts of
         Arg14, Arg28 and Asn115, and the deletion mutant lacking
         the 1-7 subsequence. Computations have been performed 
         processing with the weighted histogram method the data
         collected during folding simulations. The temperature
         is measured in units $\epsilon_0/R=151.1$ K and $C_v$ in units 
         $R=1.9855\times 10^{-3}$ Kcal mol$^{-1}$K$^{-1}$.}
\label{fig:cvplot}
\end{figure}

The specific heat profiles, displayed in Figure~\ref{fig:cvplot}, show that the
WT C5-domain and the missense mutants fold according to a cooperative, 
two-state mechanism as quantified by the van't Hoff criterion~\cite{Chan} 
determined by parameter $\kappa_2 = 2T_f \sqrt{k_B C_v(T_f)}/\Delta H_{cal}$ 
expressing the ratio between the van't Hoff and the
calorimetric enthalpies after appropriate baseline subtraction~\cite{baseline}
in energy or $C_v$ plots. 
A value of $\kappa_2$ close to unity indicates a very cooperative behavior of
the folding transition. Table~\ref{kappa2tab} summarizes the $\kappa_2$ values
along with the experimental and theoretical transition temperatures.
\begin{table}
\begin{center}
\begin{tabular}{|c|c|c|c|} \hline
Species &  $T_{exp} [K]$  & $T_{sim} [K]$ & $\kappa_2$ \\  \hline
WT      &  322.06 $\pm$ 0.75 & 322 $\pm$ 3 & 0.976  \\ \hline
Mut14   &  318.46 $\pm$ 1.44 & 322 $\pm$ 3 & 1.0    \\ \hline
Mut28   &  N.A.            & 317 $\pm$ 3 & 0.989  \\ \hline
Mut115  &  309.36 $\pm$ 0.41 & 313 $\pm$ 3 & 1.0 \\ \hline
$\Delta$ 1-7 & N. A.       & 311 $\pm$ 4 & 1.0 \\ \hline
\end{tabular}
\end{center}
\label{kappa2tab}
\caption{Experimental and simulated folding temperatures of WT domain C5
         of MyBP-C and its mutants. The last column reports the cooperativity
         parameters $\kappa_2$. N.A. = Not Available Data. }
\end{table}

Figure~\ref{fig:cvplot} also shows that a mutation on Arg14 hardly has
any effect on the thermodynamic stability of the protein as the thermogram
of this mutant is almost perfectly superposed to that of the wild-type. 
A mutation on Arg28, conversely, determines a shift of the folding 
temperature to lower values and the shift in $T_{f}$ is even larger for a
mutation on Asn115. It is worthwhile noticing that the different stability of
Mut14 and Mut28 is appreciated only using the heterogeneous model while the 
peaks of the two $C_V$ plots remain unsolved for the homogeneous G\={o}-model.
Apart for this difference in the resolution power, however the relative 
positions of the $T_f$ remains unchanged in the two models. 
The destabilizing effect of the mutation on
Asn115 appearing from our simulations is in agreement with CD and NMR spectra 
recorded by Idowu \emph{et al.}~\cite{domainC5} showing that the Asn115Lys 
mutant is unstable and largely unfolded as compared to the wild-type C5 motif. 
The same authors~\cite{domainC5} also noticed that the Arg14His mutant is 
as well folded and as stable as the wild-type which, again, is consistent 
with the good superposition of the thermograms of the wild-type and Arg14 
mutants found in our simulations. However, it is important to notice that, 
while the Asn115Lys mutant appears to be largely unfolded in mutagenesis 
experiments, a folding simulation using the G\={o}-model always ends up in 
the correct native structure and the only trace of a mutation is a shift 
in $T_f$. This is due to the fact that the G\={o}-model introduces a bias 
towards the native state so strong to override the disruptive effect of 
most mutations.

\begin{figure}
\begin{center}
\includegraphics[clip=true,keepaspectratio,scale=0.3]{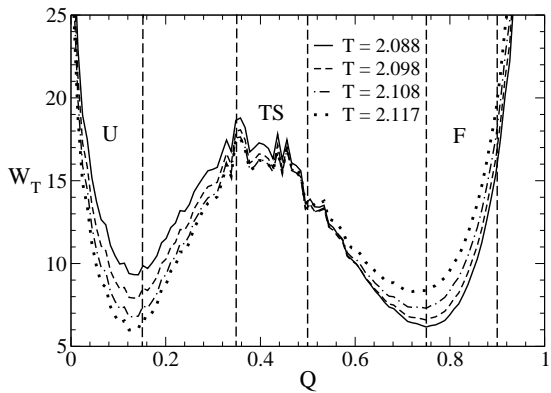}
\end{center}
\caption{Profiles of potential of mean force versus overlap around the
         folding temperature $T_f = 2.1$. The Figure shows  three 
         windows of overlap corresponding to the Unfolded ($0 < Q < 0.15$), 
         Transition State ($0.35 < Q < 0.5$) and Folded State Ensemble 
         ($0.75 < Q < 0.9$).}
\label{fig:freeQ}
\end{figure}

The 1-7 deletion mutant finally appears to be the most destabilized one
as it produces the largest shift in $T_f$. This result is also consistent 
with the observation by Idowu and coworkers that the $\Delta 1-7$ mutant 
remains largely unfolded.
The effect of the three FHC-related mutations was further investigated through
$\Phi$-value analysis. The free energy profile of the WT protein as a function
of the overlap $Q$ (fraction of native contacts) at the folding temperature 
$T_f = 2.1$, shows the typical double-well pattern of two-state folders
as illustrated in Figure~\ref{fig:freeQ}. 

\begin{figure}
\begin{center}
\includegraphics[clip=true,keepaspectratio,scale=0.35]{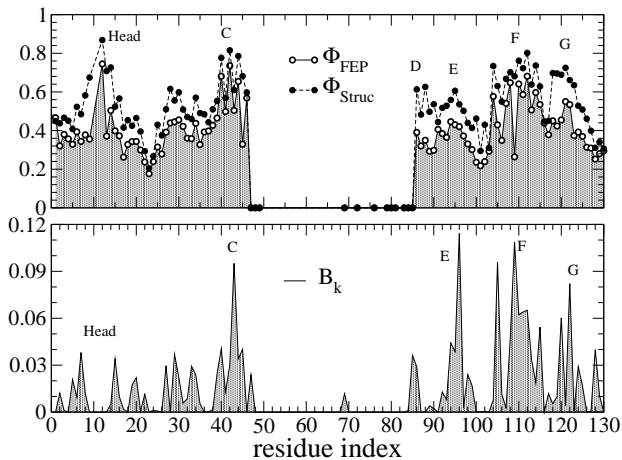}
\end{center}
\caption{Structural and perturbation $\Phi$-values (top panel) compared
         to the betweenness (bottom panel): the similarity between the 
         profiles of $\Phi$-values and betweenness suggests that the key 
         residues stabilizing the TS, derive their importance from their
         centrality in the network of contacts of the protein. The three
         profiles were computed using the conformations sampled in a run
         at folding temperature $T_f = 2.1$ in the three windows of overlap
         shown in Fig.~\ref{fig:freeQ}. "Head" refers to the N-terminal 1-17
         region.}
\label{fig:phival}
\end{figure}

The well centered values on low overlap 
correspond to the unfolded state ensemble (U), whereas the well insisting 
in the high overlap region is related to the folded state ensemble (F). 
The barrier between the two wells represents the transition state ensemble (TS). 
Conformations belonging to the F, TS and U ensembles can thus be sampled by 
the choice of three appropriate windows of the reaction coordinate $Q$ 
(Fig.~\ref{fig:phival}), and 
used for the computation of $\Phi$-values using the perturbation approach 
(Methods).

\begin{figure}
\begin{center}
\includegraphics[clip=true,keepaspectratio,scale=0.55]{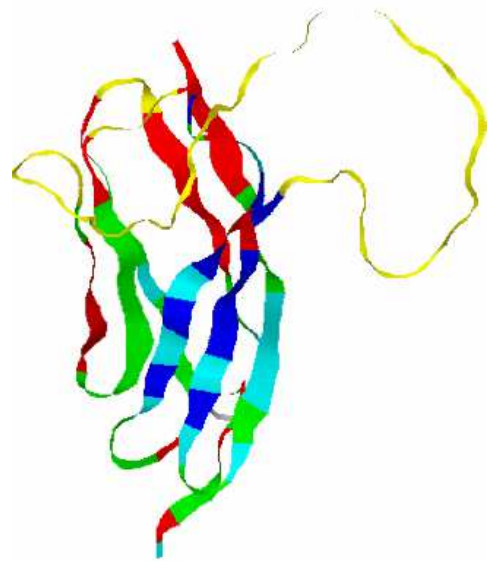}
\end{center}
\caption{Color-coded distribution of perturbation $\Phi$-values on the 
         structure of the C5 domain. Yellow: $\Phi<0.25$; Red: 
         $0.25<\Phi<0.35$; Green: $0.35<\Phi<0.45$; Cyan: $0.45<\Phi<0.55$;
         Blue: $0.55<\Phi<0.75$. The blue and cyan regions corresponding to
         the highest $\Phi$-values are concentrated on strands C, F and G.
         The BDE sheet conversely is characterized by low $\Phi$-values.}
\label{fig:struct}
\end{figure}

\begin{figure}
\begin{center}
\includegraphics[clip=true,keepaspectratio,scale=0.42]{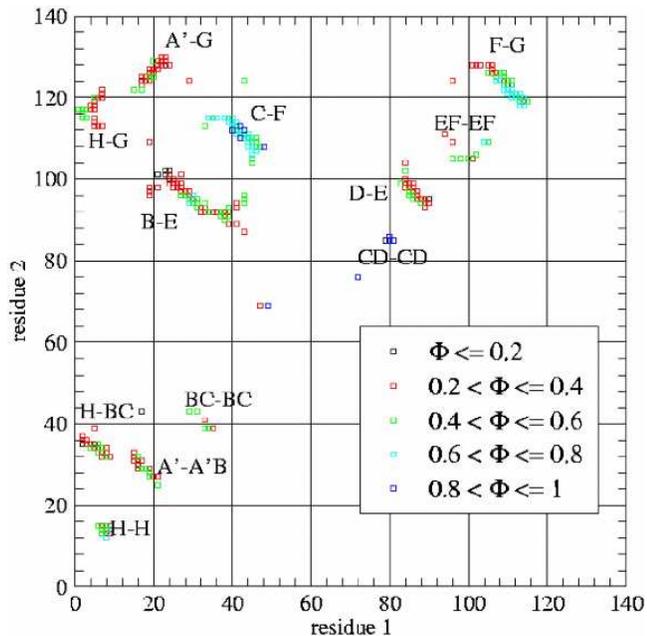}
\end{center}
\caption{Bond $\Phi$-values. The color-coded contact map shows that the
         C-F and F-G contacts feature the highest $\Phi$-values and they
         thus provide the most relevant contribution to the stability of 
         the TS. The symbol "H" designates the amino-terminal 1-17 segment.}
\label{fig:cmap}
\end{figure}

The domain C5 results to be asymmetric with respect to the distribution of
FEP $\Phi$-values, in particular the sheet containing the longest 
strands ($\beta_1$) is characterized by high $\Phi$-values while the sheet 
formed by the shorter strands ($\beta_2$) has low $\Phi$ values 
(Figure~\ref{fig:struct}). This result
is in agreement with what suggested by Clarke and coworkers~\cite{Clarke}.
In fact, the long-stranded sheet $\beta_1$, derives its high stability from
the presence of many contacts chemically corresponding to hydrogen bonds. 
The contacts more contributing to the
stability, however, are the same that stabilize the TS. 
Thus the sheet more important for the stability  ($\beta_1$) is also 
the sheet whose formation represents the rate limiting step in the folding 
kinetics.
A more detailed picture of the transition state is provided by 
Figure~\ref{fig:cmap} displaying a contact map where contact $\Phi$-values
are quantified by a color scale. The most important contacts stabilizing 
the TS are those between strands C and F and strands F and G. 
A minor contribution to the stability of the TS
is also provided by the contacts linking the central parts of 
strands B and A' and Head-Head contacts (where "Head" corresponds to the 
N-terminal 1-17 segment).
Structural $\Phi$-values offer supplementary data  
complementing the scenario supplied by the FEP analysis. The two indicators 
in fact store different information: high FEP $\Phi$-values identify 
those residues whose mutation most destabilizes the TS. Whereas 
high structural $\Phi$-values characterize residues 
in a native-like conformation in the TS regardless of whether they 
actively stabilize the structure or they were passively driven in a 
correct native-like conformation by the rearrangement of neighboring 
regions of the molecule. This pattern was also observed in an 
experimental study on a fibronectin-like domain reported in 
Ref.~\cite{fibro}.
The peculiar features of the two indicators thus explain why structural
$\Phi$-values are systematically higher than FEP $\Phi$-values. The difference
between structural and perturbation $\Phi$-values yields a profile whose peaks
identify the residues passively driven in a native-like conformation in the TS:
this residues are mainly located on strand D and at the boundary between 
strands F and G. It can also be noticed that  Pro12 plays an active role in 
stabilizing the N-terminal region in a native-like conformation in the TS, 
while His8 and Gly13 are just passively placed in the correct position.

The analysis of betweenness (Fig.~\ref{fig:phival}) shows that this 
parameter, representing the fraction of minimal paths passing through a given 
residue, correlates well with the $\Phi$-values, in agreement
with Ref.~\cite{Vendruscolo}. 
In particular, the betweenness confirms the importance of 
strands C, F and G but it differs from $\Phi$-values in two important regions 
of the protein. Strand E is characterized by a high betweenness as it is the
central strand of the BDE sheet  and it probably acts as a bridge between
 strands B and D: the importance of this strand may thus be higher than it
appears from $\Phi$-values alone. Conversely, the N-terminal region of the 
protein is characterized by a low betweenness so that it appears to be 
 weakly connected with the other parts of the molecule.   

It is instructive to discuss the positions of the 3 FHC-related mutations
within the two $\beta$-sheets of the C5 domain. In fact, Asn115 which lies
on the sheet characterized by the highest FEP $\Phi$-values, when mutated
to Lys, is known to completely disrupt the native structure of this protein.
On the other hand, Arg14 and Arg28 are located on the sheet with low
FEP $\Phi$-values and their mutation does not significantly affect the 
thermodynamic stability.

\begin{figure}
\begin{center}
\includegraphics[clip=true,keepaspectratio,scale=0.35]{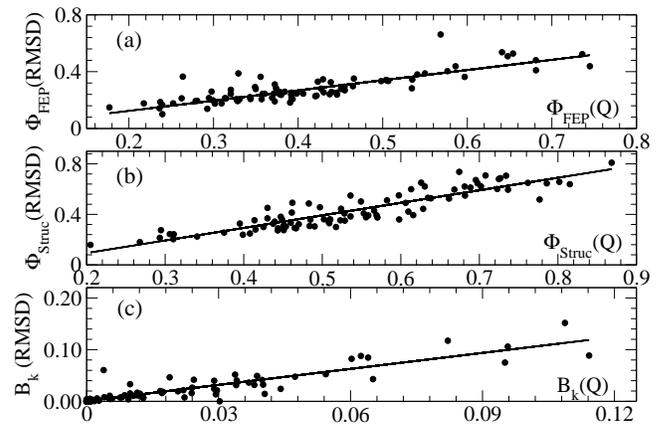}
\end{center}
\caption{Linear regression of parameters computed using
the Kabsch RMSD and the fraction of native contacts $Q$. 
(a) Free energy Perturbative $\Phi$-values computed using Kabsch RMSD
    reaction coordinate versus the same quantity computed using the overlap $Q$.    Correlation coefficient $r = 0.8$. 
(b) Structural $\Phi$-values (Kabsch RMSD)
    versus $\Phi_{Struc}$  ($Q$).  Correlation coefficient $r = 0.99$.
(c) Betweenness (Kabsch RMSD)
    versus $B_k$  ($Q$).  Correlation coefficient $r = 0.90$.}
\label{fig:corr}
\end{figure}

As a final remark, we tested the sensitivity of betweenness and structural
and perturbative $\Phi$-values on the choice of the reaction coordinate. 
The computation protocol for these parameters has been repeated by using the 
Kabsch RMSD \cite{Kabsch} as a collective coordinate for sampling of F, TS and U ensemble 
structures. The existence of linear relations (Fig.~\ref{fig:corr}), 
with correlations coefficient greater than $0.8$,  
between the parameters computed using either the 
Kabsch RMSD or the fraction of native contacts $Q$ indicates that 
the sampling of the F, TS and U conformations is equivalent for the two
methods. Thus the information conveyed by the parameters $\Phi_{FEP}$, 
$\Phi_{struct}$, $B_k$ is statistically significative because not strongly 
dependent on the reaction coordinates used to compute them.  

\subsection{The CD loop}
As already mentioned in Section~\ref{Intro}, one of the most interesting 
features of the cardiac isoform of the C5 domain of MyBP-C is the presence
of a 28 residue long insert that makes the CD loop significantly longer 
(residue 47 to 85) than the corresponding loop of the skeletal isoforms. 
Our simulations show that the CD loop is extremely mobile as it is 
involved in very few native contacts. 

\begin{figure}[h!]
\begin{center}
\includegraphics[clip=true,keepaspectratio,scale=0.35]{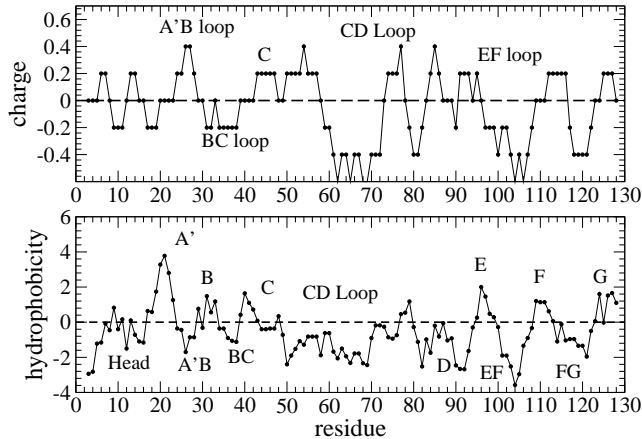}
\end{center}
\caption{Charge (top panel) and hydrophobicity (bottom panel) profiles
         along the protein sequence. Each value is an average over
         a 5-residue long window shifted along the polypeptide chain.
         Hydrophobicity was computed using the Kyte-Doolittle scale; 
         regions with a positive value are hydrophobic. The CD loop appears
         to have a high density of negative charge and is scarcely 
         hydrophobic. "Head" designates the 1-17 segment of the protein.}
\label{fig:bioinf}
\end{figure}

This high mobility of the CD loop is an extremely important feature of the
C5 domain because it is responsible for a folding temperature significantly
lower than that of most Ig domains~\cite{domainC5}. 
The reasons for this unusual mobility
can be clarified through a simple sequence analysis. First of all, we 
analyzed the hydrophobicity along the amino acid sequence using the Kyte 
and Doolittle scale~\cite{Kyte}. The hydrophobicity of each residue was
calculated by sliding a 5-residue long window over the protein sequence
and assigning to the central residue the average hydrophobicity computed 
over the entire window. A similar approach was employed for the charge
where Glu and Asp residues contribute -1, Lys and Arg contribute +1 and 
the other residues are regarded as being neutral at physiological pH. 
The profiles in Figure~\ref{fig:bioinf} show that all the loops,
and in particular the CD loop, feature an excess of 
negative charge and are less hydrophobic than the regions corresponding to
the $\beta$-strands.    
An analysis of the amino acid distribution along the
chain, reveals a high concentration of charged residues  in the CD        
loop where the number of Glu (4) and Asp (5) residues exceeds the number of
Lys (3) and Arg (2) residues. A high concentration of Glu is also found
in the EF loop. A remarkable feature is also the high concentration of
Pro in the N-terminal 1-17 region, in the BC and in the CD loops. 
In summary, the CD loop is characterized by a high concentration of the 
residues identified by Garner \emph{et al.}~\cite{Dunker,Garner,Romero}  
as strong determinants of local disorder. Moreover, 
Uversky~\cite{Uversky1,Uversky2} showed that the combination of low mean 
hydrophobicity and relatively high net charge represents a prerequisite 
for the absence of compact structure in proteins under physiological 
conditions. In particular, it was shown that  the charge-hydrophobicity 
phase-space can be divided into two regions by the empirical separatrix 
line of equation:
\begin{equation} 
\langle H \rangle_{boundary} = \frac{\langle R \rangle + 1.151}{2.785} 
\label{separatrix} 
\end{equation}
where $H$ refers to hydrophobicity and $R$ to the charge. The proteins located
below this line in the phase-space are likely to be unfolded in physiologic
conditions whereas those lying above the separatrix do fold in a compact, 
globular conformation. In order to test this issue, the mean hydrophobicity and
the mean net charge were computed for the CD loop, for the portion of the 
protein sequence not including this loop and for the whole protein sequence.
Following Uversky, in these calculations the hydrophobicity of individual 
residues was normalized to a scale of 0 to 1, and the mean hydrophobicity is 
computed as the sum of the normalized hydrophobicities  divided by the number 
of residues in the protein segment under examination. A similar approach was 
used for the computation of the mean charge. Table~\ref{table1} shows that 
the CD loop is located in the natively unfolded region of the phase space. The
full sequence of the C5 domain is closer to the boundary of the natively 
folded region that, however, it cannot reach due to the effect of the CD loop
in the computation of the average charge and hydrophobicity. Finally the 
portion of the polypeptide sequence not including the CD loop is even closer to
the boundary but it still relies in the natively unfolded region due to the
features of the minor loops.
\begin{table}
\begin{center}
\begin{tabular}{|l|c|c|c|}
\hline
Molecular Region & $\langle R \rangle$ & $\langle H \rangle$ &
$\langle H \rangle_{Bound}$  \\
\hline
Full Protein     & 0.546               & 0.417 & 0.609  \\
\hline
CD Loop          & 0.507               & 0.343 & 0.595  \\
\hline
Outside CD Loop  & 0.563               & 0.450 & 0.615   \\
\hline
\end{tabular}
\end{center}
\caption{Positions in the charge-hydrophobicity phase-space  of three
         moleculer regions of Domain C5 of MyBP-C.
         The CD loop, the full C5 domain and the portion of the protein
         excluding the CD loop all lie in the natively unfolded region of
         the phase-space below the separatrix (Eq~\ref{separatrix}).}
\label{table1}
\end{table}
The properties of the CD sequence were further studied by  analyzing
the average number of native contacts per amino-acid as suggested 
in Ref.~\cite{Galzitskaya}. According to this approach, natively
unfolded proteins are supposed to form a number of native interactions
insufficient to compensate for the loss of conformational entropy, hence 
their necessity to couple folding with specific ligand binding. As a 
consequence, natively unfolded proteins are expected to feature an 
average number of contacts lower than that of globular proteins. 

Natively unfolded and globular proteins can also be discriminated 
through
a set of 20 artificial parameters designed through Monte Carlo 
maximization~\cite{Galzitskaya} of the scoring function 
$\mbox{Score} = (\langle X_f \rangle - \langle X_u \rangle) /
\sqrt{S_{f}^{2} + S_{u}^{2}}$
where $\langle X_f \rangle$ and $\langle X_u \rangle$ are the mean values
of the adjustable parameters in two training dataset of folded and
natively unfolded proteins and $S_{f}$ and $S_{u}$ are the corresponding
standard deviations. 
 
\begin{figure}[h!]
\begin{center}
\includegraphics[clip=true,keepaspectratio,scale=0.35]{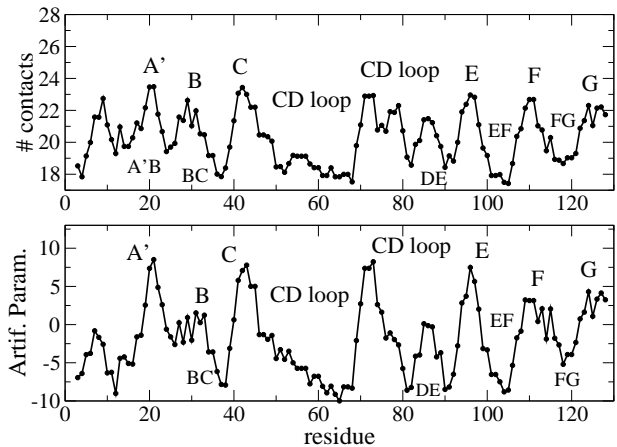}
\end{center}
\caption{Profiles of average number of contacts (top) and artificial
         parameters (bottom) along the sequence of C5 domain. The
         computation was performed using the parameters listed in Table I
         of Ref.~\cite{Galzitskaya}. The loops are characterized 
         by  lower values of both indicators with respect to the strand
         subsequences.}
\label{fig:bioinf2}
\end{figure}

We generated the profiles of the average number of contacts per
residue and of the artificial parameters by shifting a 5-residue
window along the protein sequence and assigning the window average to
the central residue, so that data in Fig.~\ref{fig:bioinf2} can be 
compared 
to the charge and hydrophobicity plots of Fig.~\ref{fig:bioinf}, determined 
with the same procedure. The profiles of the average number of contacts
and of the artificial parameters in Figure~\ref{fig:bioinf2}, show that both
indicators are effective in discriminating $\beta$-strands and unstructured
loops, the latter being characterized by much lower values of the
parameters. 
These results thus suggest that the CD loop of
domain C5 of MyBP-C should be classified as a \emph{natively unfolded sequence}
\emph{i.e.} it is  protein fragment that lacks a stable structure even 
in physiological conditions. 
A common feature of natively unfolded proteins is that their 
folding is usually associated with the binding to a specific ligand. 
It could thus be suggested that the long CD loop that appears structureless 
when the C5
domain is dissected from MyBP-C, might actually be well folded \emph{in vivo}
due to a close interaction with a specific ligand. 
This hypothesis is further supported by the experimental finding that 
cardiac MyBP-C co-purifies with 
the Calmodulin class-II (CaM-II)-like Kinase~\cite{Hartzell,Gautel,Redwood2} 
so that the folding of the CD loop may be accompanied by docking with this 
enzyme.

\section{Discussion and Conclusions}

The involvement of MyBP-C in Familial Hypertrophic Cardiomyopathy motivated
a study of the folding of domain C5 through equilibrium MD simulations
to gain insight into the role of the three FHC-related mutations:
Asn115Lys, Arg14His and Arg28His. 
As a member of the Immunoglobulin family, domain C5 lends itself 
to be reasonably modeled through a G\={o}-like force field~\cite{Clarke}. 
We  assessed the thermodynamics impact of a mutation through the entity of the shift
in the folding temperature. Our results show that, among the three 
FHC-related mutations
we examined, Asn115Lys determines the largest decrease in $T_f$, 
in agreement with the NMR spectra recorded by
Idowu~\cite{domainC5} signalling absence of structure even at low 
temperature. Conversely, the $T_f$ shift induced by the Arg28His mutation 
is modest,  while the protein destabilization induced by Arg14His 
is negligible as its thermogram is almost perfectly superposable 
to that of the WT. 
This finding suggests that the latter two mutants have very 
little effect on protein stability and their pathological role must be sought 
elsewhere. Both mutations Arg14His and Arg28His imply the removal of three contacts and 
their impact in the G\={o}-like  approach could be partially resolved only through the 
introduction of heterogeneous energetic couplings suggesting the
opportunity of a more refined analysis. 

Further insights in the role of Arg14 located in the N-terminus of the C5 
domain, were attained through the study of the $\Delta 1-7$ deletion mutant.
The significant decrease in $T_f$ of the truncated domain indicates that the
N-terminal region with its 10-residue long insert typical of the cardiac 
isoform, is not just a linker between the C4 and C5 domains, but it gives an 
important contribution to the stability. However, the low betweenness of the 
N-terminal residues indicates that they may be involved in contacts forming 
a subgraph only weakly connected to the core of the contact network. 
It is thus possible 
that the N-terminal contacts do appear only when the C5 domain is dissected 
from the rest of the protein and that the natural role of this section is
more related to the binding with domain C8 complementing the negatively 
charged CFGA' surface~\cite{domainC5}.

This potential role of the FHC-related mutations is confirmed by the analysis 
of $\Phi$-values that appear to be significantly higher in the CFGA' sheet where
Asn115 is located, than in the BDE sheet that including Arg268.
 
A final issue considered in the present work, is a analysis of the CD
loop responsible for the low stability of the C5 domain as compared with
other Ig domains.
The charge unbalance and the low hydrophobicity of the CD loop, accompanied
by a low average number of native contacts and a low value of the artificial
parameters introduced by Galzitskaya~\cite{Galzitskaya}, is a clear indication 
that the C5 domain of MyBP-C can be considered a \emph{natively unfolded} 
protein \emph{i.e.} a protein that lacks a compact, globular structure under
physiological conditions~\cite{Dunker,Garner,Romero,Uversky1,Uversky2}. 
Therefore the role of the CD loop of the C5 domain of the cardiac isoform 
of MyBP-C, must be reconsidered within the framework of the peculiar
properties of natively unfolded proteins. As the cardiac MyBP-C co-purifies
with the Calmodulin class-II (CaM-II) like Kinase, the CD loop may represent
an SH3 domain recognition region~\cite{Hartzell,Gautel,Redwood2}. 
In experiments and simulations performed on 
the C5 domain alone, the CD loop due to its high mobility, destabilizes the 
protein, lowering its folding temperature. In vivo, however, the CD loop may
fold upon binding with the CaM-II-like Kinase, so that the thermodynamic
stability and the folding temperature of the protein may be similar to those
of the other Ig domains. 
Our results also suggest that the cardiac 
C5 domain might regulate the activity of the CaM-II-like Kinase whose
docking may trigger the folding of the CD loop.
In such a case, the C5 domain may be not only a structural component of the 
Moolman-Smook collar (Fig.~\ref{ms-model}) but it also may play an important 
role in the regulation of muscular contraction. 


\end{document}